\documentclass[10pt,twocolumn,journal]{IEEEtran}

\usepackage{lettrine}
\usepackage{amsmath}
\usepackage{cases}
\usepackage{bm}
\usepackage{txfonts}
\usepackage{amssymb}
\usepackage{cite}
\usepackage{graphicx}
\usepackage{psfrag}
\usepackage{url}
\usepackage{stfloats}
\usepackage{algorithm}
\usepackage{algorithmic}
\usepackage{multirow}
\usepackage{color}
\usepackage{romannum}
\usepackage{lipsum}
\usepackage{stfloats}
\usepackage[utf8]{inputenc}
\usepackage[skip=2pt,font=scriptsize]{caption}
\usepackage[skip=2pt,font=scriptsize]{subcaption}
\usepackage{tabularx}
\usepackage{booktabs}

\ifCLASSINFOpdf

\else
\fi

\begin{document}
\pagenumbering{arabic}
\title{Real-Time and Security-Aware Precoding in RIS-Empowered Multi-User Wireless Networks}

\author{Abuzar B. M. Adam,~\IEEEmembership{Member,~IEEE},  Mohamed Amine Ouamri, Mohammed Saleh Ali Muthanna,~\IEEEmembership{Member,~IEEE}, Xingwang Li,~\IEEEmembership{Senior Member,~IEEE}, Mohammed A. M. Elhassan, Ammar Muthanna,~\IEEEmembership{Senior Member,~IEEE}
\thanks{\IEEEauthorrefmark{1}Corresponding author: Abuzar B. M. Adam  (abuzar@cqupt.edu.cn)}
\thanks{A. B. M. Adam is with the Interdisciplinary Centre for Security, Reliability and Trust (SnT), University of Luxembourg, 1855, Luxembourg City, Luxembourg, (email: abuzar.babikir@uni.lu).}
\thanks{M. A. Ouamri is with Computer communication University Grenoble Alpes, CNRS, Grenoble INP, LIG Laboratory, DRAKKAR Teams, 38000 Grenoble, France.}
\thanks{ M. S. A. Muthanna is with School of Communications and Information Engineering, Chongqing University of Posts and Telecommunications, Chongqing,
P. R. China, 400065.(e-mail: st103044@student.spbu.ru)}
\thanks{Xingwang Li is with the School of Physics and Electronic Information Engineering, Henan Polytechnic University, Jiaozuo 454000, China(e-mail: lixingwangbupt@gmail.com)}
\thanks{Mohammed A. M. Elhassan is with School of Mathematics and Computer Science, Zhejiang Normal University (e-mail: Mohammedac29@zjnu.edu.cn).}
\thanks{ A.Muthanna is with the Peoples' Friendship University of Russia (RUDN University) 6 Miklukho-Maklaya, 117198 Moscow, Russia; muthanna.asa@spbgut.ru}
}

\maketitle

\begin{abstract}
In this letter, we propose a deep-unfolding-based framework (DUNet) to maximize the secrecy rate in reconfigurable intelligent surface (RIS) empowered multi-user wireless networks. To tailor DUNet, first we relax the problem, decouple it into beamforming and phase shift subproblems, and propose an alternative optimization (AO) based solution for the relaxed problem. Second, we apply Karush-Kuhn-Tucker (KKT) conditions to obtain a closed-form solutions for the beamforming and the phase shift. Using deep-unfolding mechanism, we transform the closed-form solutions into a deep learning model (i.e., DUNet) that achieves a comparable performance to that of AO in terms of accuracy and about 25.6 times faster.

\end{abstract}


\section{Introduction}

\lettrine[lines=2]{T}{o} provide coverage to the inaccessible spots and enhancing the connectivity in the sixth generation (6G) networks, reconfigurable intelligent surfaces (RISs) can be used to establish line-of-sight (LoS) links\cite{md1,md2,bb1}. Due to its passive nature, RIS are energy-efficient and flexible for installation \cite{bb2}. However, RIS-assisted communications entail security issues due to the passive eavesdroppers which require secrecy-aware designs as well as difficulties in acquiring channel state information (CSI) \cite{bb3}. Besides, RIS-assisted communications are complex and often associated with highly nonconvex problems \cite{bb3}. Different studies investigated some of these problems \cite{bb4,bb6}. However, the proposed solutions in these studies are iterative, slow, and may not meet the targets of the six generation (6G) networks in terms of real-time inference.

Recently, deep learning approaches have been widely used to handle different problems in wireless systems \cite{md3,md4,md6}. Some studies considered deep learning to design solutions for RIS-assisted networks \cite{md7,md9}. Most of the studies in the literature have considered data-driven deep learning models which require large data set for training. Hence, to decrease the required amount of data while achieving near optimal performance, deep-unfolding mechanism can be used to design highly efficient models.

In this work, we propose a deep-unfolding based framework (DUNet) to optimize the secrecy rate in multiuser RIS-assisted communication. We formulate the optimization problem as a secrecy rate maximization and propose a solution for the relaxed version of the problem. Next, we apply the optimality conditions to design DUNet based on the closed-form expression of the beamforming and the phase shift. The proposed deep learning model achieves performance comparable to that of the numerical solution and about 25.6 times faster.

\section{System Model And Problem Formulation}
\begin{figure}[ht]
\centerline{\includegraphics[width=2.5in]{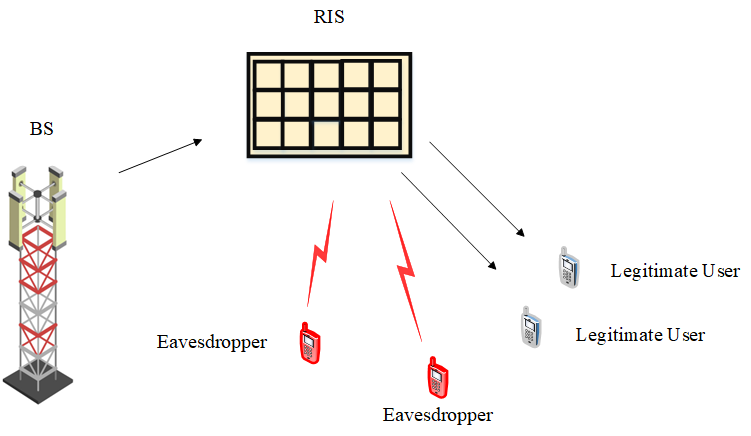}}
\caption{RIS-empowered multiuser network.}
\label{sys}
\end{figure}
We consider a downlink multiuser RIS network as shown in Fig. \ref{sys}, where we have a base station (BS) and communicates with $K$ legitimate users while $E$ ground eavesdroppers are attempting to wiretap $K$ communications. The BS, legitimate users, and eavesdroppers are equipped multiple antennas denoted as $N_B, N_K, $ and $N_E$ formulating uniform linear arrays (ULA). The RIS is equipped with $J$ elements. The coordinates of UAV, legitimate user $k$, eavesdropper $e$, and the RIS are respectively given as $C_B = \left(x_B,y_B,H_B\right)^T$, $C_k = \left(x_k,y_k\right)^T$, $C_e = \left(x_e,y_e\right)^T$, and $C_R = \left(x_R,y_R,z_R\right)^T$.

Assuming that the LoS link is blocked and the received signals at the users are scattering components. Hence, using Rayleigh fading model, the channel modelling between the BS and $k$ and $e$ is given as
\begin{equation}\label{eqn1}
{{\bf{h}}_{B,k}} = \sqrt {\rho d_{B,k}^{ - \alpha }} \left( {\sqrt {\frac{{{{\rm{R}}_{B,k}}}}{{{{\rm{R}}_{B,k}} + 1}}} {\bf{\tilde h}}_{B,k}^{LoS} + \sqrt {\frac{1}{{{{\rm{R}}_{B,k}} + 1}}} {\bf{\tilde h}}_{B,k}^{NLoS}} \right) \in {\mathbb{C}^{{N_B} \times {N_K}}},
\end{equation}
\begin{equation}\label{eqn2}
{{\bf{h}}_{B,e}} = \sqrt {\rho d_{B,e}^{ - \alpha }} \left( {\sqrt {\frac{{{{\rm{R}}_{B,k}}}}{{{{\rm{R}}_{B,k}} + 1}}} {\bf{\tilde h}}_{B,e}^{LoS} + \sqrt {\frac{1}{{{{\rm{R}}_{B,e}} + 1}}} {\bf{\tilde h}}_{B,e}^{NLoS}} \right) \in {\mathbb{C}^{{N_B} \times {N_E}}},
\end{equation}
\begin{equation}\label{eqn3}
{{\bf{h}}_{B,R}} = \sqrt {\rho d_{B,R}^{ - \kappa}} \left( {\sqrt {\frac{{{{\rm{R}}_{B,R}}}}{{{{\rm{R}}_{B,R}} + 1}}} {\bf{\tilde h}}_{B,R}^{LoS} + \sqrt {\frac{1}{{{{\rm{R}}_{B,R}} + 1}}} {\bf{\tilde h}}_{B,R}^{NLoS}} \right) \in {\mathbb{C}^{{J} \times {N_B}}},
\end{equation}

where $\rho$ is the channel power gain at the reference distance $d_0 = 1$ m, ${d_{B,i}} = \sqrt {\left\| {{C_B} - {C_i}} \right\|} ,i \in \left\{ {k,e,R} \right\}$ is the distance between the BS and the users, and $\alpha$ and $\kappa$ are the pathloss exponents. $\rm R_{B,i}$ is the Rician factor between the BS and users. ${{\bf{\tilde h}}_{B,i}^{NLoS}},i \in \left\{ {k,e,R} \right\}$ are assumed to be independent and identically distributed variables following circularly symmetric complex Gaussian distribution with zero mean and unit variance. ${{\bf{\tilde h}}_{B,k}^{LoS}}$ and ${{\bf{\tilde h}}_{B,e}^{LoS}}$ are given as follows
\begin{equation}\label{eqn3a}
{\bf{\tilde h}}_{B,k}^{LoS} = {\bf{a}}_{{N_B}}^T{{\bf{a}}_{{N_k}}},
\end{equation}
\begin{equation}\label{eqn3b}
{\bf{\tilde h}}_{B,e}^{LoS} = {\bf{a}}_{{N_B}}^T{{\bf{a}}_{{N_e}}},
\end{equation}
where

\begin{equation*}
\begin{array}{*{20}{l}}
{{{\bf{a}}_{{N_B}}} = \left[ {\begin{array}{*{20}{l}}
{1,\exp \left( {j\frac{{2\pi {d_B}}}{\lambda }\left( {\cos {\varpi _{B,i}}\sin {\zeta _{B,i}}} \right)} \right),...,}\\
{\exp \left( {j\frac{{2\pi {d_B}}}{\lambda }\left( {{N_B} - 1} \right)\left( {\cos {\varpi _{B,i}}\sin {\zeta _{B,i}}} \right)} \right)}
\end{array}} \right],i\left\{ {k,e} \right\},}\\
\begin{array}{l}
{{\bf{a}}_{{N_k}}} = \left[ {\begin{array}{*{20}{l}}
{1,\exp \left( {j\frac{{2\pi {d_k}}}{\lambda }\left( {\cos {\phi _{B,k}}\sin {\vartheta _{B,k}}} \right)} \right),...,}\\
{\exp \left( {j\frac{{2\pi {d_k}}}{\lambda }\left( {{N_K} - 1} \right)\left( {\cos {\phi _{B,k}}\sin {\vartheta _{B,k}}} \right)} \right)}
\end{array}} \right],\\
{{\bf{a}}_{{N_e}}} = \left[ {\begin{array}{*{20}{l}}
{1,\exp \left( {j\frac{{2\pi {d_e}}}{\lambda }\left( {\cos {\phi _{B,e}}\sin {\vartheta _{B,e}}} \right)} \right),...,}\\
{\exp \left( {j\frac{{2\pi {d_e}}}{\lambda }\left( {{N_E} - 1} \right)\left( {\cos {\phi _{B,e}}\sin {\vartheta _{B,e}}} \right)} \right)}
\end{array}} \right]
\end{array}
\end{array}
\end{equation*}

where $d_B, d_k,$ and $d_e$ are the antenna elements separation distance for the case of BS, legitimate user $k$ , and eavesdropper $e$, respectively. ${\zeta _{B,i}}$ and ${\varpi _{B,i}}$ are the azimuth and elevation angles of departure. ${\vartheta _{B,i}}$ and ${\phi _{B,i}}$ represent are the azimuth and elevation angles of arrival. The LoS channel between the ${{\bf{\tilde h}}_{B,R}^{LoS}}$ is given as follows
\begin{equation}\label{eqn3b}
{\bf{\tilde h}}_{B,R}^{LoS} = {\bf a}_J^T{{\bf a}_{{N_B}}},
\end{equation}
The array responses ${\bf a}_J$ and ${{\bf a}_{{N_B}}}$ are defined as
\begin{equation}\label{eqn4}
\begin{aligned}
{{\bf a}_J} &= \left[ {1,...,\exp \left( { - j\frac{{2\pi }}{\lambda }{d_x}\left( {{J_x} - 1} \right)\sin \phi \cos \varphi } \right)} \right]\\
 &\otimes \left[ {1,...,\exp \left( { - j\frac{{2\pi }}{\lambda }{d_z}\left( {{J_z} - 1} \right)\cos \phi } \right)} \right]
\end{aligned}
\end{equation}
\begin{equation}\label{eqn5}
{{\bf{a}}_{{N_B}}} = \left[ \begin{array}{l}
1,\exp \left( { - j\frac{{2\pi }}{\lambda }{d_{{B}}}\cos \theta } \right),...,\\
\exp \left( { - j\frac{{2\pi }}{\lambda }{d_{{B}}}\left( {{N_B} - 1} \right)\cos \theta } \right)
\end{array} \right],
\end{equation}
where $d_x$ and $d_z$ represent RIS elements separation distance along x-axis and z-axis, respectively. $J_x$ and $J_z$ are the RIS elements along the x-axis and z-axis, respectively. $\sin \phi \cos \varphi  = \frac{{{x_R} - {x_B}}}{{{d_{B,R}}}}$ and $\cos \phi  = \frac{{{H_B} - {z_R}}}{{{d_{B,R}}}}$ with $\varphi$ and $\phi$ respectively represent the azimuth angle of arrival and the elevation angle of arrival. $\lambda$ is the wavelength. $\theta$ is the angle of departure. The channels between the users and the RIS contain both LoS and NLoS. Using Rician channel modeling, these channels are given as
\begin{equation}\label{eqn6}
{{\bf{h}}_{R,i}} = \sqrt {\rho d_{R,i}^{ - \kappa }} \left( {\sqrt {\frac{{\rm{R}}_{R,i}}{{{\rm{R}}_{R,i} + 1}}} {\bf{\tilde h}}_{R,i}^{LoS} + \sqrt {\frac{1}{{{\rm{R}}_{R,i} + 1}}} {\bf{\tilde h}}_{R,i}^{NLoS}} \right),i \in \left\{ {k,e} \right\}
\end{equation}
where $d_{R,i}$ is the distance between the RIS and the user $i$ and $\kappa$ is the pathloss exponent. ${\bf{\tilde h}}_{R,k}^{NLoS} \in {\mathbb{C}^{J \times {N_K}}}$ and ${\bf{\tilde h}}_{R,e}^{NLoS} \in {\mathbb{C}^{J \times {N_E}}}$ respectively denote the NLoS link between the RIS and the user, RIS and the eavesdropper, which are modeled as complex Gaussian distributed with zero mean and unit variance. ${\bf{\tilde h}}_{R,i}^{LoS}$ is the LoS link between the user and the RIS can be expressed as
\begin{equation}\label{eqn7}
\begin{aligned}
{\bf{\tilde h}}_{R,i}^{LoS} &= \left[ {1,...,\exp \left( { - j\frac{{2\pi }}{\lambda }{d_x}\left( {{J_x} - 1} \right)\sin {\phi _i}\cos {\varphi _i}} \right)} \right]\\
 &\otimes \left[ {1,...,\exp \left( { - j\frac{{2\pi }}{\lambda }{d_z}\left( {{J_z} - 1} \right)\cos {\phi _i}} \right)} \right]
\end{aligned}
\end{equation}
where $\varphi_i$ and $\phi_i$ respectively represent the azimuth angle of arrival and the elevation angle of arrival of the RIS and the user link. The transmit signal at the BS can be expressed as
\begin{equation}\label{eqn8}
{\rm{x}} = {\bf{Ws}} + {\bf{v}}{\rm{,}}
\end{equation}
where ${\bf{W}} = \left[ {{{\bf{w}}_1}{\rm{,}}{{\bf{w}}_2},...,{{\bf{w}}_K}} \right]$ with ${{\bf{w}}_k} \in {\mathbb{C}^{{N_B} \times {N_K}}}$ is the precoding matrix, ${\bf{s}}$ is the information signal with $\mathbb{E}\left( {{{\bf{s}}_k}{\bf{s}}_k^H} \right) = {\bf{I}}_{N_K}$, and ${\mathbf{v}}$ is the artificial noise vector due to the hardware imperfections where ${\bf{v}} \sim {\cal C}{\cal N}\left( {0,\upsilon {\rm{diag}}\left\{ {\sum\limits_{i = 1}^K {{{\bf{w}}_i}{\bf{w}}_i^H} } \right\}} \right)$ and $\upsilon  \in \left[ {0,1} \right]$ is the proportionality coefficient to characterize the hardware impairments. Let ${\mathbf{\Theta}}  = {\rm{diag}}\left( {\exp \left( {j{\omega _1}} \right),..,\exp \left( {j{\omega _m}} \right),...,\exp \left( {j{\omega _J}} \right)} \right)$ with ${\omega _m}$ represents the phase shift of the element $m$. The received signal can be given as
\begin{equation}\label{eqn9}
\begin{aligned}
{{{{\bf{d}}_k}}} & {{= \left( {{\bf{h}}_{B,k}^H + {{\bf{h}}_{R,k}^H}\Theta {{\bf{h}}_{B,R}}} \right)\left( {{{\bf{w}}_k}{s_k} + {\bf{v}}} \right)}}\\
 &{{+ \left( {{\bf{h}}_{B,k}^H + {{\bf{h}}_{R,k}^H}\Theta {{\bf{h}}_{B,R}}} \right)\sum\limits_{j \in K/k} {{{\bf{w}}_j}{s_j}} {\rm{ + }}{{\bf{z}}_k}}}
\end{aligned}
\end{equation}
{{where ${z_k} \sim {\cal C}{\cal N}\left( {0,\sigma _k^2{{\bf{I}}_{N_K}}} \right)$ is the additive white Gaussian noise (AWGN). Similarly, the signal of the eavesdropper $e$ can be expressed as ${{\bf{d}}_e} = \left( {{\bf{h}}_{B,e}^H + {{\bf{h}}_{R,e}^H}\Theta {{\bf{h}}_{B,R}}} \right)\left( {{{\bf{w}}_k}{s_k} + {\bf{v}}} \right) + \left( {{\bf{h}}_{B,e}^H + {{\bf{h}}_{R,e}^H}\Theta {{\bf{h}}_{B,R}}} \right)\sum\limits_{j \in K/k} {{{\bf{w}}_j}{s_j}} {\rm{ + }}{{\bf{z}}_e}$ with ${z_e} \sim {\cal C}{\cal N}\left( {0,\sigma _e^2{{\bf{I}}_{N_E}}} \right)$}}. Let the estimated cascaded channel matrix for the user $k$ be denoted as ${{\bf{G}}_k} = {\left[ {{{\bf{g}}_1},...,{{\bf{g}}_{{N_K}}}} \right]^T} \in {\mathbb{C}^{{N_K} \times {N_B}J}},$ with ${{\bf{g}}_{{N_K}}} = {\left[ {{\bf{g}}_{1,{N_K}}^T,...,{\bf{g}}_{{N_B},{N_K}}^T} \right]^T}$ denotes the estimated cascaded channel vector at the receiving antenna ${N_K}$.
The received signal in \eqref{eqn9} can be rewritten for the case of the user $k$ as follows
\begin{equation}\label{eqn9a}
\begin{array}{l}
{{\bf{d}}_k} = \left( {{\bf{h}}_{B,k}^H + {{\bf{G}}_k}\Theta {{\bf{1}}_F}} \right)\left( {{{\bf{w}}_k}{s_k} + {\bf{v}}} \right)\\
 + \left( {{\bf{h}}_{B,k}^H + {{\bf{G}}_k}\Theta {{\bf{1}}_F}} \right)\sum\limits_{j \in K/k} {{{\bf{w}}_j}{s_j}} {\rm{ + }}{{\bf{z}}_k},
\end{array}
\end{equation}
where ${{\bf{1}}_F} = {\left[ {{\bf{1}}_1^T,...,{\bf{1}}_{{N_B}}^T} \right]^T}$ with  where the elements of the column $N_B$ are ones and all other columns are zeros. Thus, ${{\bf{G}}_k}\Theta {{\bf{1}}_F} \in {^{{N_K} \times {N_B}}}$. The signal-to-interference-plus-noise ratio (SINR) ${\gamma _k}$ is given as below
\begin{equation}\label{eqn10}
{{{\gamma _k} = \frac{{{{\left| {\left( {{\bf{h}}_{B,k}^H + {{\bf{G}}_k}\Theta{{\bf{1}}_F}} \right){{\bf{w}}_k}} \right|}^2}}}{{\left[ {\begin{array}{*{20}{l}}
{\underbrace {\sum\limits_{j \in K/k} {{{\left| {\left( {{\bf{h}}_{B,k}^H + {{\bf{G}}_k}\Theta{{\bf{1}}_F}} \right){{\bf{w}}_j}} \right|}^2}\left( {\left( {1 + \upsilon } \right)\upsilon {\rm{diag}}\left( {{{\bf{w}}_j}{\bf{w}}_j^H} \right)} \right)} }_{{\rm{interfernece}}}}\\
{ + \underbrace {{\mho_k}}_{{\rm{distortion}}} + \underbrace {\left( {1 + \upsilon } \right)\sigma _k^2}_{{\rm{noise}}}}
\end{array}} \right]}}}}
\end{equation}
where ${\mho_k}$ is the signal distortion due to the hardware imperfections and given as
\begin{equation*}
\begin{aligned}
{{\mho_k}}  & {{= \left( {{\bf{h}}_{B,k}^H + {{\bf{G}}_k}\Theta{{\bf{1}}_F}} \right)\left( {\upsilon {{\bf{w}}_k}{\bf{w}}_k^H + \left( {1 + \upsilon } \right)\upsilon {\rm{diag}}\left( {{{\bf{w}}_k}{\bf{w}}_k^H} \right)} \right)}}\\
 &{{\times \left( {{{{\left( {{{\bf{G}}_k}\Theta {{\bf{1}}_F}} \right)}^H}} + {{\bf{h}}_{B,k}}} \right)}}
\end{aligned}
\end{equation*}
The hardware imperfections have impact on the interference and the noise levels as it can be seen in the first and third terms in the denominator of $\gamma_k$.

Similarly, we can define the SINR for the eavesdropper $e$
\begin{equation}\label{eqn12}
{{{\gamma _e} = \frac{{{{\left| {\left( {{\bf{h}}_{B,e}^H + {{\bf{G}}_e}\Theta{{\bf{1}}_F}} \right){{\bf{w}}_k}} \right|}^2}}}{{\left[ \begin{array}{l}
\sum\limits_{j \in K/k} {{{\left| {\left( {{\bf{h}}_{B,e}^H + {{\bf{G}}_e}\Theta{{\bf{1}}_F}} \right){{\bf{w}}_j}} \right|}^2}\left( {\left( {1 + \upsilon } \right)\upsilon {\rm{diag}}\left( {{{\bf{w}}_j}{\bf{w}}_j^H} \right)} \right)} \\
 + {\mho_e}+ \left( {1 + \upsilon } \right)\sigma _e^2
\end{array} \right]}},}}
\end{equation}
where $\mho_e$ is the signal distortion due to the hardware imperfections and given as
\begin{equation*}
\begin{aligned}
{{{\mho _e}}} & {{= \left( {{\bf{h}}_{B,e}^H + {{\bf{G}}_e}\Theta{{\bf{1}}_F}} \right)\left( {\upsilon {{\bf{w}}_k}{\bf{w}}_k^H + \left( {1 + \upsilon } \right)\upsilon {\rm{diag}}\left( {{{\bf{w}}_k}{\bf{w}}_k^H} \right)} \right)}}\\
 &{{\times \left( {\left({{\bf{G}}_e}\Theta{{\bf{1}}_F}\right)^H + {{\bf{h}}_{B,e}}} \right)}}
\end{aligned}
\end{equation*}
Channel state information (CSI) of the eavesdropper cannot be obtained. Therefore, using estimated channel error, we can define the channel coefficients of the eavesdropper as
\begin{equation}\label{eqn13}
\begin{array}{l}
{{\bf{h}}_{B,e}} = {{{\bf{\hat h}}}_{B,e}} + \Delta {{\bf{h}}_{B,e}},\\
{{\bf{G}}_e} = {{{\bf{\hat G}}}_e} + \Delta {{\bf{G}}_e},
\end{array}
\end{equation}
where $\Delta {{\bf{h}}_{B,e}} = {{\iota}_{e,1}}{\mathfrak{g}_{e,1}} \sim {\cal C}{\cal N}\left( {0,{\iota_{e,1}}{{\bf{I}}_{{N_B} \times {N_E}}}} \right)$ and $\Delta {{\bf{G}}_e} = {\iota_{e,2}}{\mathfrak{g}_{e,2}} \sim {\cal C}{\cal N}\left( {0,{\iota_{e,2}}{{\bf{I}}_{{N_B} \times {N_E}}}} \right)$. ${\iota_{e,1}}$ and ${\iota_{e,2}}$ are constants measuring the level of channel uncertainties.
The achievable rates is defined as follows
\begin{equation}\label{eqn14}
{r_k} = {\log _2}\left( {1 + {\gamma _k}} \right)
\end{equation}
and
\begin{equation}\label{eqn15}
{r_e} = {\log _2}\left( {1 + {\gamma _e}} \right)
\end{equation}
The secrecy rate is defined as
\begin{equation}\label{eqn16}
R_k^{\sec } = {r_k} - {r_e}
\end{equation}
We aim at maximizing the secrecy rate through optimizing the beamforming and the phase shift. The optimization problem is formulated as follows
\begin{subequations}\label{eqn17:main}
\begin{align}
&\mathop {\max }\limits_{{\bf{w}}, \Theta} && \sum\limits_{k \in {\cal K}} {R_k^{\sec }}   &   & \tag{\ref{eqn17:main}} \\
& \text{s.t.}&& {{\sum\limits_{k=1}^{K}{\|{{\bf{w}}_k}\|^2}  \preceq {P_B},}}\label{eqn17:b} \\
&             &&{{\|{{\bf{w}}_k}\|^2 \succeq 0,\label{eqn17:c}}}\\
&             &&{r_k} \ge \varepsilon_k ,\label{eqn17:d} \\
&             &&\mathop {\max }\limits_{\Delta {{\bf{h}}_{B,e}},\Delta {{\bf{G}}_e}}{r_e} \le \varepsilon_e ,\label{eqn17:e} \\
&             &&\left| {\exp \left( {j{\omega _m}} \right)} \right| = 1.
\end{align}
\end{subequations}
Constraint \eqref{eqn17:b} is the power budget constraint with $P_B$ represents the total transmit power. Constraints \eqref{eqn17:d} and \eqref{eqn17:e} are the secrecy constraints which ensures that the rate of the legitimate user is above the limit $\varepsilon_k$ and the rate of the eavesdropper is below the limit $\varepsilon_e$; where $0 < {\varepsilon _e} < {\varepsilon _k}$. Since constraints \eqref{eqn17:d} and \eqref{eqn17:e} represent lower and upper bounds, problem \eqref{eqn17:main} is nonconvex. In the following sections, we provide a traditional solution for problem \eqref{eqn17:main} and based on this solution, we design a deep learning framework using deep unfolding techniques.

\section{Alternative Optimization (AO) Method}
To facilitate derivation, {we follow the semi-definite relaxation (SDR) procedure in \cite{mod211} to reformulate} the problem in \eqref{eqn17:main} as
\begin{subequations}\label{eqn18:main}
\begin{align}
&\mathop {\max }\limits_{{\bf{Q}}, \Theta} && \sum\limits_{k \in {\cal K}} {R_k^{\sec }}   &   & \tag{\ref{eqn18:main}} \\
& \text{s.t.}&&\sum\limits_{k \in {\cal K}} {{\rm{Tr}}\left( {{{\bf{Q}}_k}} \right)}  \preceq {P_B},\label{eqn18:b} \\
&             &&{{{\bf{Q}}_k}} \succeq 0,\label{eqn18:c}\\
&             &&{r_k} \succeq \varepsilon_k ,\label{eqn18:d} \\
&             &&\mathop {\max }\limits_{\Delta {{\bf{h}}_{B,e}},\Delta {{\bf{G}}_e}}{r_e} \preceq \varepsilon_e ,\label{eqn18:e} \\
&             &&\left| {\exp \left( {j{\omega _m}} \right)} \right| = 1,\label{eqn18:f} \\
&             &&{\rm{Rank}}\left( {{{\bf{Q}}_k}} \right) = 1.\label{eqn18:g}
\end{align}
\end{subequations}
{Constraint \eqref{eqn18:e} still intractable, we apply Bernstein-Type inequality \cite{abm1} to transform \eqref{eqn18:e}. Towards this, we define the slack variables $\mathfrak{n}$, $\mathfrak{a}$, and $\mathfrak{b}$. Thus, constraint\eqref{eqn18:e} can rewritten as}
\begin{equation}\label{eqn20}
\mathfrak{g}_e^H{{\bf{\Phi }}_e}{\mathfrak{g}_e} + 2{\mathop{\rm Re}\nolimits} \left\{ {{{\bf{B}}^H}\mathfrak{g}{_e}} \right\} + \hat c \le 0
\end{equation}
where
\begin{equation*}
{\mathfrak{g}_e} = {\left[ {\begin{array}{*{20}{c}}
{\mathfrak{g}_{e,1}^H}&{\mathfrak{g}_{e,2}^T}
\end{array}} \right]^H},
\end{equation*}
\begin{equation*}
{{\bf{\Phi }}_e} = \left[ {\begin{array}{*{20}{c}}
{\begin{array}{*{20}{c}}
{\iota_{e,1}^2\Omega }\\
{{\iota_{e,2}}{\iota_{e,2}}\left( {\Omega  \otimes {\left({\Theta{\bf{1}}_F}\right)^*}} \right)}
\end{array}}&{\begin{array}{*{20}{c}}
{{\iota_{e,2}}{\iota_{e,2}}\left( {\Omega  \otimes {\left({\Theta{\bf{1}}_F}\right)^T}} \right)}\\
{\iota_{e,2}^2\left( {\Omega  \otimes {{\bf{A}}^T}} \right)}
\end{array}}
\end{array}} \right],
\end{equation*}
\begin{equation*}
\Omega  = {\rm{Tr}}\left( \begin{array}{l}
{\eta _e}{{\bf{Q}}_k} - \sum\limits_{j \in {\cal K}/k} {{{\bf{Q}}_j}\left( {1 + \upsilon } \right)\upsilon {\rm{diag}}\left( {{{\bf{Q}}_j}} \right)} \\
 - \upsilon {{\bf{Q}}_k}\left( {1 + \upsilon } \right)\upsilon {\rm{diag}}\left( {{{\bf{Q}}_k}} \right)
\end{array} \right),
\end{equation*}
\begin{equation*}
 \resizebox{0.45\textwidth}{!}{${\bf{B}} = {\left[ {\begin{array}{*{20}{c}}
{{\iota_{e,1}}\left( {{\bf{\hat h}}_{B,e}^H + {{{\bf{\hat G}}}_e{\Theta{\bf{1}}_F}}} \right)\Omega }&{{\iota_{e,2}}{\rm{ve}}{{\rm{c}}^T}\left( {\left( {{\bf{\hat h}}_{B,e}^H + {{{\bf{\hat G}}}_e{\Theta{\bf{1}}_F}}} \right)\left({\Theta{\bf{1}}_F}\right)^H\Omega } \right)}
\end{array}} \right]^H},$}
\end{equation*}
\begin{equation*}
\hat c = \left( {{\bf{\hat h}}_{B,e}^H + {{{\bf{\hat G}}}_e{\Theta{\bf{1}}_F}}} \right)\Omega \left( {{{{\bf{\hat h}}}_{B,e}} + {\left({\bf{\hat G}}_e{\Theta{\bf{1}}_F}\right)}^H} \right) + \left( {1 + \upsilon } \right)\sigma _e^2,
\end{equation*}
with ${\bf{A}} = \left({\Theta{\bf{1}}_F}\right)\left({\Theta{\bf{1}}_F}\right)^H$. Then, using Bernstein-Type inequality, we can express \eqref{eqn18:e} as in the following
\begin{equation}\label{eqn21}
\left\{ {\begin{array}{*{20}{c}}
{{\rm{Tr}}\left( {{{\bf{\Phi }}_e}} \right) - \sqrt {2\ln \left( {{1 \mathord{\left/\mathfrak{n}
 {\vphantom {1 }} \right.
 \kern-\nulldelimiterspace} }} \right)}  - \ln \left( {{1 \mathord{\left/\mathfrak{n}
 {\vphantom {1 }} \right.
 \kern-\nulldelimiterspace} }} \right)\mathfrak{b} + \hat c \le 0,}\\
{{{\left\| {\begin{array}{*{20}{c}}
{{\rm{vec}}\left( {{{\bf{\Phi }}_e}} \right)}\\
{\sqrt 2 {\bf{B}}}
\end{array}} \right\|}_2} \le \mathfrak{a},}\\
{\mathfrak{b}{\bf{I}} + {{\bf{\Phi }}_e} \ge 0, \mathfrak{b}\ge 0,}
\end{array}} \right.
\end{equation}

Based on the above procedure, the optimization problem in \eqref{eqn18:main} can be approximated as
\begin{subequations}\label{eqn22:main}
\begin{align}
&\mathop {\max }\limits_{{\bf{Q}}, \Theta, \mathfrak{a}, \mathfrak{b}} && \sum\limits_{k \in {\cal K}} {R_k^{\sec }}  &   & \tag{\ref{eqn22:main}} \\
& \text{s.t.}&&\eqref{eqn18:b},\eqref{eqn18:c}, \eqref{eqn18:d},\eqref{eqn18:f},\eqref{eqn18:g}, \eqref{eqn21},\notag\\
&             && \mathfrak{b}\geq 0,\label{eqn22:a}
\end{align}
\end{subequations}
The optimization problem can be divided into beamforming subproblem and phase shift subproblem. The two subproblems can be jointly solved. The beamforming subproblem is given as below
\begin{subequations}\label{eqn23:main}
\begin{align}
&\mathop {\max }\limits_{{\bf{Q}}} && \sum\limits_{k \in {\cal K}} {R_k^{\sec }} &   & \tag{\ref{eqn23:main}} \\
& \text{s.t.}&&\eqref{eqn18:b},\eqref{eqn18:c}, \eqref{eqn18:d}, \eqref{eqn18:g}, \eqref{eqn22:a},\notag\\
&            &&\begin{aligned}
\left( {\iota _{e,1}^2 + \iota _{e,2}^2J} \right){\rm{Tr}}\left( \Omega  \right) &- \sqrt {2\ln \left( {{1 \mathord{\left/\mathfrak{n}
 {\vphantom {1 }} \right.
 \kern-\nulldelimiterspace} }} \right)}\\
&  - \ln \left( {{1 \mathord{\left/\mathfrak{n}
 {\vphantom {1 }} \right.
 \kern-\nulldelimiterspace} }} \right)\mathfrak{b} + \hat c \le 0,
\end{aligned}\label{eqn23:a}\\
&           &&{\left\| {\begin{array}{*{20}{c}}
{\left( {\iota_{e,1}^2 + \iota_{e,2}^2J} \right){\rm{vec}}\left( \Omega  \right)}\\
{\sqrt {2\left( {\iota_{e,1}^2 + \iota_{e,2}^2J} \right)} \Omega \left( {{{{\bf{\hat h}}}_{B,e}} + {\left({\bf{\hat G}}_e{\Theta{\bf{1}}_F}\right)}^H} \right)}
\end{array}} \right\|_2} \le \mathfrak{a},\label{eqn23:b}\\
&          &&{\mathfrak{b}\bf{I}} + \left( {\iota_{e,1}^2 + \iota_{e,2}^2J} \right)\Omega  \ge 0,\label{eqn23:c}
\end{align}
\end{subequations}
Using SDR \cite{bb7} and the techniques in \cite{abm1},  the convex problem \eqref{eqn23:main} can be efficiently handled using CVX. The phase shift subproblem is given as follows
\begin{subequations}\label{eqn24:main}
\begin{align}
&\mathop {\max }\limits_{{\bf{A}}, \mathfrak{a},\mathfrak{b}} &&\sum\limits_{k \in {\cal K}} {R_k^{\sec }}  &   & \tag{\ref{eqn24:main}} \\
& \text{s.t.}&& \eqref{eqn18:f}, \eqref{eqn21},
\end{align}
\end{subequations}
Similarly, problem \eqref{eqn24:main} is convex and can be tackled using the same procedure as in the case of the beamforming subproblem. Algorithm 1 illustrates the main steps of solving \eqref{eqn17:main}.
\begin{algorithm}[H]
\label{alg}
\caption {AO for solving problem \eqref{eqn17:main}}
\begin{algorithmic}[1]
\renewcommand{\algorithmicrequire}{\textbf{Initialization:}}
\REQUIRE $\mathbf{Q}^{\left(0\right)}$, ${\Theta}^{\left( 0 \right)}$, $\iota_{e,1}$, $\iota_{e,2}$, $\mathfrak{a}^{\left( 0 \right)}$, $\mathfrak{b}^{\left( 0 \right)}, \upsilon$, $\mathfrak{n}, \varepsilon_k, \varepsilon_e$, $\tau = 0$, and $\epsilon$
\WHILE{$\left| \sum\limits_{k \in {\cal K}} {R_k^{\sec }}{\left(\tau\right)} - \sum\limits_{k \in {\cal K}} {R_k^{\sec }}{\left(\tau - 1\right)}\right| \ge \epsilon$}
\STATE $\tau = \tau + 1$
\STATE Solve \eqref{eqn24:main} to obtain ${\Theta}^{\left( \tau \right)}$, $\mathfrak{a}^{\left( \tau \right)}$, and $\mathfrak{b}^{\left( \tau \right)}$ with $\mathbf{Q}^{\left(\tau - 1\right)}$, ${\Theta}^{\left( \tau-1 \right)}$, $\mathfrak{a}^{\left( \tau - 1 \right)}$ and $\mathfrak{b}^{\left( \tau - 1\right)}$.
\STATE Update $\mathbf{Q}_k^{\left(\tau\right)}$ by solving \eqref{eqn23:main} with $\mathbf{Q}^{\left(\tau -1\right)}$, ${\Theta}^{\left( \tau \right)}$, $\mathfrak{a}^{\left( \tau \right)}$ and $\mathfrak{b}^{\left( \tau \right)}$
\ENDWHILE
\end{algorithmic}
\end{algorithm}
\begin{figure*}
\centering
  \includegraphics[width=11cm,height=4cm]{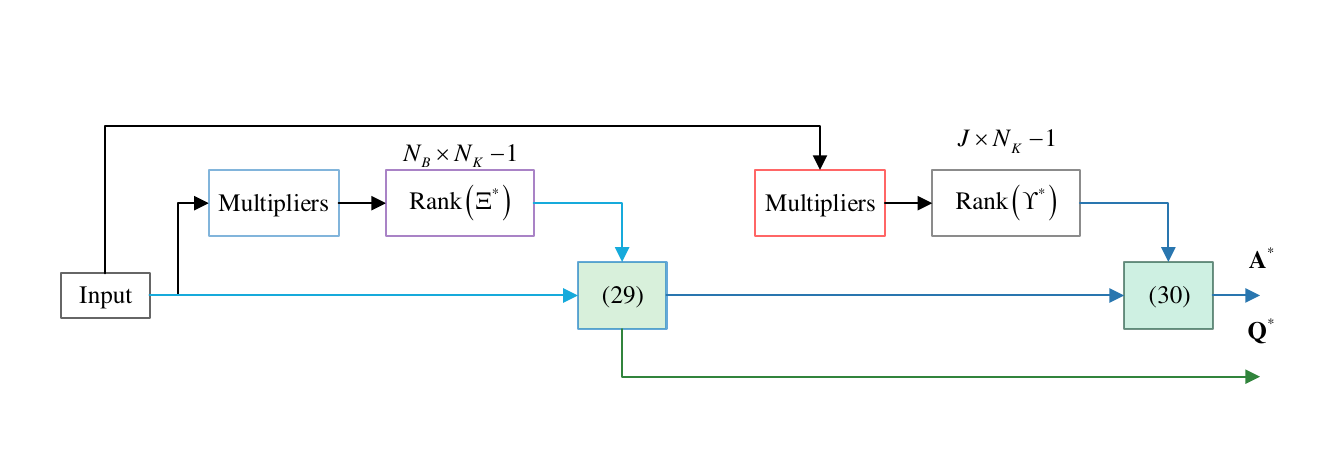}
  \caption{Structure of the proposed DUNet.}
  \label{model}
\end{figure*}

\section{Deep-Unfolding based Framework (DUNet)}
In this section, we discuss the details of the proposed deep-unfolding-based framework (DUNet). The proposed DUNet utilizes the proposed solution AO as grounds and convert it via deep-unfolding into a multilayer neural network \cite{md10}. Below, we go through the procedure of obtaining the closed-form expressions while considering the rank of the solution for the beamforming and the unit-modulus constrain for the case of the phase shift. The architecture of the proposed is illustrated in Fig. \ref{model}.

The numerical solution in Section III can obtain the optimal solution by performing eigenvalue decomposition if the obtain solution admits Rank-one. However, the occurrence of ${\rm Rank}\left( {{\bf{Q}}_i^*} \right) > 1$ is possible. Hence, the proposed DUNet is designed in such a way to convey a solution that admits Rank-one.

We can employ the tightness of the relaxed problems \eqref{eqn23:main} and \eqref{eqn24:main} to design the proposed DUNet. To show that the problems are convex with respect to the optimization variables and satisfy slater's qualifications, Karush-Kuhn-Tucker (KKT) conditions are necessary and sufficient conditions \cite{bb7} for the solution of the problems. We construct the following Lagrangian function
\begin{equation}\label{eqn25}
\resizebox{0.70\hsize}{!}{$
\begin{aligned}
{{{\cal L}}} & {{= \sum\limits_{k \in K} {R_k^{\sec }}  + \mu \left( {{P_B} - \sum\limits_{k \in K} {{\rm{Tr}}\left( {{{\bf{Q}}_k}} \right)} } \right) - {\rm{Tr}}\left( {{\Xi _k}{{\bf{Q}}_k}} \right)}}\\
& {{- {\rm{Tr}}\left( {{{\bf{C}}_2}{{\bf{D}}_{{{\bf{C}}_1}}}} \right) - {\rm{Tr}}\left( {{{\bf{C}}_2}{{\bf{D}}_{{{\bf{C}}_2}}}} \right)}}\\
& {{- {\rm{Tr}}\left( {{{\bf{C}}_3}{{\bf{D}}_{{{\bf{C}}_3}}}} \right) - {\rm{Tr}}\left( {{\bf{A}}\Upsilon } \right) + {\mathbf{\psi}}}}
\end{aligned}$}
\end{equation}
{where ${\mathbf{D}_{\mathbf{C}_1}}, {\mathbf{D}_{\mathbf{C}_2}}$, and ${\mathbf{D}_{\mathbf{C}_3}}$, are the set of Lagrange multipliers associated with \eqref{eqn23:a}, \eqref{eqn23:b}, and \eqref{eqn23:c} with ${\bf{A}}^*$ representing the optimal solution of \eqref{eqn24:main}.} $\mathbf{\psi}$ represents the set of variables associated with other constraints. The KKT conditions are given as
\begin{subequations}\label{eqn26:main}
\begin{equation}
\resizebox{0.30\hsize}{!}{${\nabla_{\bf{Q}}}{\cal L} = 0,{\nabla_{\bf{A}}}{\cal L} = 0,$}\label{eqn26:a}
\end{equation}
\begin{equation}
\resizebox{0.50\hsize}{!}{$\mu  \ge 0,{\rm B}_k^*,{\Upsilon ^*},{\bf{D}}_{{{\bf{C}}_1}}^*,{\bf{D}}_{{{\bf{C}}_2}}^*,{\bf{D}}_{{\bf{C}}3}^*,{\bf{D}}_{{{\bf{C}}_4}}^* \succeq 0,$}\label{eqn26:b}
\end{equation}
\begin{equation}
\Xi _k^*{\bf{Q}}_k^* = 0,\label{eqn26:c}
\end{equation}
\begin{equation}
{{\bf{A}}^*}{\Upsilon ^*} = 0,\label{eqn26:d}
\end{equation}
\end{subequations}
Applying KKT conditions yields the following closed-form expression
\begin{equation}\label{eqn27}
{\bf{Q}}_k^* = {\left[ {\frac{1}{{\Xi _k^* - {\mu ^*}}} + {{\bf{S}}^*}} \right]^ + },
\end{equation}
where
\begin{equation*}
\resizebox{0.80\hsize}{!}{$
\begin{aligned}
{{\bf{S}}^*} &= \frac{1}{{\ln 2{{{\bf{\bar D}}}^*}}} - \frac{{{{\bf{I}}^*} - \left( {1 + \upsilon } \right)\sigma _i^2}}{{{\bf{\mathord{\buildrel{\lower3pt\hbox{$\scriptscriptstyle\smile$}}
\over N} }}}},\\
{{{{{\bf{\bar D}}}^*}}} & {{= {\bf{D}}_{{{\bf{C}}_1}}^* \left( {{\rm vec}\left( {1 + \upsilon } \right){\bf{A}} - {\bf{A}}\left( {1 + \upsilon } \right)\upsilon {\rm diag}\left( {{{\bf{I}}_{{N_B} \times {N_K}}}} \right)} \right)+ {\bf{D}}_{{{\bf{C}}_2}}^*}}\\
& {{+ {\bf{D}}_{{\bf{C}}3}^* + {\bf{D}}_{{{\bf{C}}_4}}^*,}}\\
{{{{\bf{I}}^*}}} & {{= {\rm{Tr}}\left(\left( {{\bf{\hat h}}_{B,k}^H + {{{\bf{\hat G}}}_k}\Theta {{\bf{1}}_F}} \right){\bf{\Phi}}_k\left( {{\bf{\hat h}}_{B,k} + \left({{{\bf{\hat G}}}_k}\Theta {{\bf{1}}_F}\right)^H} \right) \right),}}\\
{{{\bf{\mathord{\buildrel{\lower3pt\hbox{$\scriptscriptstyle\smile$}}
\over N} }}}} & {{= {\rm{Tr}}\left(\left( {{\bf{\hat h}}_{B,k}^H + {{{\bf{\hat G}}}_k}\Theta {{\bf{1}}_F}} \right){\bf{A}}\left( {{\bf{\hat h}}_{B,k} + \left({{{\bf{\hat G}}}_k}\Theta {{\bf{1}}_F}\right)^H} \right)\right),}}
\end{aligned}$}
\end{equation*}
It can be observed that ${{\bf{\bar D}}^*},{{\bf{I}}^*},{\bf{\mathord{\buildrel{\lower3pt\hbox{$\scriptscriptstyle\smile$}}
\over N} }},$ and ${{\bf{S}}^*}$ are semi-definite and ${\rm Rank}\left( {{{\bf{S}}^*}} \right) = {N_B} \times {N_K}$ whereas ${\rm Rank}\left( {{\Xi ^*}} \right) = {N_B} \times {N_K}$ or  ${N_B} \times {N_K} - 1$. Hence, we choose ${\rm Rank}\left( {{\Xi ^*}} \right) = {N_B} \times {N_K} - 1$ to guarantee that ${\bf{Q}}_k^*$ lies in the null space of  $\Xi _k^*$ and thus ${\rm Rank}\left( {{\bf{Q}}_k^*} \right) = 1$. With similar analysis, we obtain the closed-form expression for ${\bf{A}}$ as in \eqref{eqn28} on top of next page.
\begin{figure*}
\begin{equation}\label{eqn28}
{{
\resizebox{0.70\hsize}{!}{${{\bf{A}}^*} = {\left[ {\frac{{\sqrt {{{\left( {2\left( {1 + \upsilon } \right)\left( {\sigma _i^2 - \sigma _e^2} \right) + {\bf{\hat h}}_{B,e}^H{{{\bf{\hat G}}}_e}} {\bf{Q}}_k^*\right)}^2} - \Sigma }  - 1}}{{\left( \begin{array}{l}
2\left( {1 + \upsilon } \right)\left( {\sigma _i^2 - \sigma _e^2} \right)
\left( \begin{array}{l}
\sum\limits_{j \in K/k} {{\rm{Tr}}\left( {\left( {{\bf{\hat h}}_{B,k}^H + {{{\bf{\hat G}}}_k}\Theta {{\bf{1}}_F}} \right){\bf{Q}}_j^*\left( {\left( {1 + \upsilon } \right)\upsilon {\rm{diag}}\left( {{\bf{Q}}_j^*} \right)} \right)\left( {{{{\bf{\hat h}}}_{B,k}} + {{\left( {{{{\bf{\hat G}}}_k}\Theta {{\bf{1}}_F}} \right)}^H}} \right)} \right)} \\
 + \left( {\upsilon {\bf{Q}}_k^* + \left( {1 + \upsilon } \right){\rm{diag}}\left( {{\bf{Q}}_k^*} \right)} \right)
\end{array} \right)
\end{array} \right)}}} \right]^ + }$}}}
\end{equation}
where
\begin{equation*}
{{\resizebox{0.70\hsize}{!}{$\Xi  = 4\left( {1 + \frac{{\left( {{\bf{\hat h}}_{B,k}^H + {{{\bf{\hat G}}}_k}\Theta {{\bf{1}}_F}} \right){\bf{Q}}_k^*\left( {{{{\bf{\hat h}}}_{B,k}} + {{\left( {{{{\bf{\hat G}}}_k}\Theta {{\bf{1}}_F}} \right)}^H}} \right)}}{{\left( {\begin{array}{*{20}{l}}
\begin{array}{l}
{\bf{D}}_{{{\bf{C}}_1}}^*\left( {{\rm{vec}}\left( {1 + \upsilon } \right){\bf{Q}}_k^* - {\rm{vec}}\left( {1 + \upsilon } \right)\upsilon {\rm{diag}}\left( {{\bf{Q}}_k^*} \right)} \right)
 + {\bf{D}}_{{{\bf{C}}_2}}^*\left( {\iota_{e,2}^2\left( {\Omega  \otimes {{\bf{I}}_J}} \right)} \right)
\end{array}\\
\begin{array}{l}
 + {\bf{D}}_{{\bf{C}}3}^*\left( {{\rm{vec}}\left( {\iota_{e,2}^2\left( {\Omega  \otimes {{\bf{I}}_J}} \right)} \right)} \right)
 + {\bf{D}}_{{{\bf{C}}_4}}^*\left( {\iota_{e,2}^2{\rm{ve}}{{\rm{c}}^T}\left( {\Omega {{{\bf{\hat G}}}_e}{{\bf{I}}_{J \times {N_K}}}} \right)} \right) + {\Upsilon ^*}
\end{array}
\end{array}} \right)}}} \right)$}}}
\end{equation*}
\end{figure*}

where ${\rm Rank}\left({\Upsilon ^*}\right) = J \times N_K - 1$ to guarantee that ${\bf{A}}^*$ lies in null space of ${\Upsilon ^*}$, and thus ${\rm Rank}\left({\bf{A}}^*\right) = 1$. Using the solution in \eqref{eqn27} and \eqref{eqn28}, we build the basic layer of DUNet as in Fig. \ref{model}. To decrease the number of layer while enhancing the accuracy, we use the incremental learning mechanism in \cite{bb8} to train DUNet.

\section{Simulation Results}
The default number of legitimate users is set to 8 and the number of eavesdroppers is set to 4; unless stated otherwise. {In all cases, the users are divided into number of groups equal to the number of the eavesdroppers. For example, 8 legitimate users are divided in 4 groups where each eavesdropper wiretap two a group}. BS is equipped with four antennas while all users are equipped with two antennas for each.  $\kappa  = \alpha  = 3.2$,  ${d_x} = {d_z} = {\lambda  \mathord{\left/
 {\vphantom {\lambda  4}} \right.
 \kern-\nulldelimiterspace} 4}$,  ${d_B} = {\lambda  \mathord{\left/
 {\vphantom {\lambda  2}} \right.
 \kern-\nulldelimiterspace} 2}$, and $R = 4$. Noise power is set to -80 dBm, and $\epsilon = 0.001$.

%
%

\begin{figure}[htbp]
\centerline{\includegraphics[width=3in]{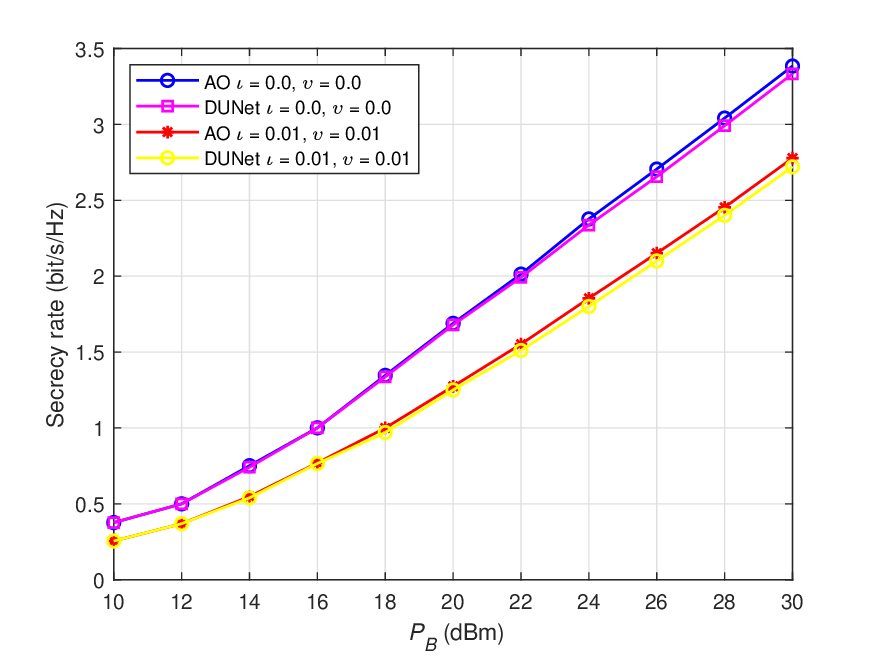}}
\caption{Secrecy rate versus $P_B$.}
\label{fig4}
\end{figure}

From Fig. \ref{fig4} we can observe that the proposed DUNet achieves similar performance to that of AO in terms of secrecy rate for different values of the BS power $P_B$. Increasing the value of $\iota$ and $\upsilon$ decreases the secrecy because higher power consumption is required to compensate the signal loss due to hardware impairments and channel errors. Additionally, the prediction accuracy of the proposed DUNet is impacted by the levels of signal distortion and channel errors due to the statistical nature of these parameters. Nevertheless, DUNet achieves 99.61\% performance of AO and as it will be shown in Fig. \ref{fig7}, DUNet is drastically faster compared to AO.

\begin{figure}[htbp]
\centerline{\includegraphics[width=3in]{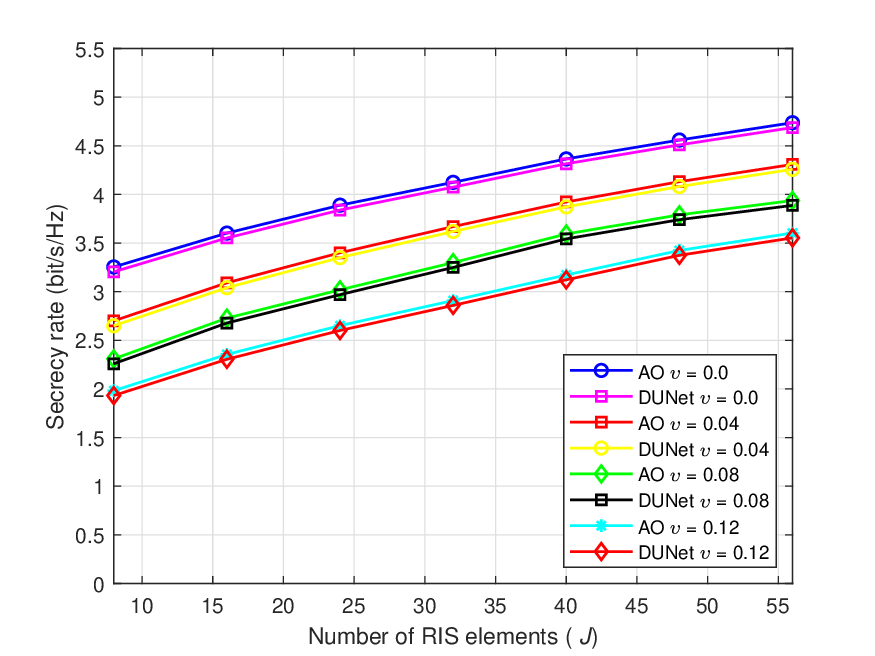}}
\caption{Achievable secrecy rate versus RIS elements.}
\label{fig5}
\end{figure}

Fig. \ref{fig5} illustrates the performance comparison between AO and DUNet for under the impact of hardware level $\upsilon$ and for different number RIS elements. $K = 12,E = 4,$ and ${P_B} = 30$ dBm. The secrecy rate increases with the increase of the number of RIS elements and that is because the signal from RIS becomes dominant and a wider margin to perform secure phase shift optimization is obtained. Moreover, the increasing of $\upsilon$ leads to decreasing in the secrecy rate due to the need for higher power to compensate. DUNet still robust the drastic increase of $\upsilon$ and has closer performance to that of AO.

\begin{figure}[htbp]
\centerline{\includegraphics[width=3in]{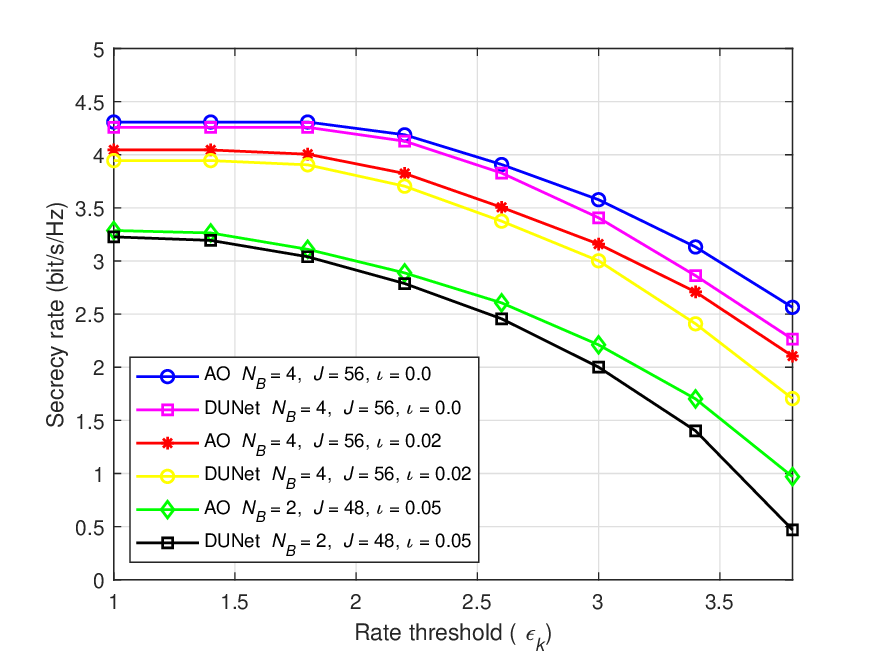}}
\caption{Secrecy rate versus rate threshold  $\varepsilon_k$.}
\label{fig6}
\end{figure}

Fixing the ${\varepsilon _e}$ at 1, we plot the secrecy rate versus the rate threshold ${\varepsilon _k}$ while considering the impact of number of antennas, the number of RIS elements, and the channel uncertainty. This is depicted in Fig \ref{fig6}. $\upsilon  = 0.02,K = 12,$ and $E = 4$. The value of ${\varepsilon _k}$ determines the secrecy and confidentiality of the message of the user $k$. We can observe from the figure that the increase of ${\varepsilon _k}$ leads to the decreasing of the secrecy rate and that is because the increase in the gap ${\varepsilon _k} - {\varepsilon _e}$ incurs more power consumption the meet the secrecy constraints. The increasing of $N_B$ and $J$ leads better performance by improving the channel gain, enhancing the signal strength, and providing margin to optimize the beamforming and phase shift. The performance gap between the AO and DUNet increases for higher ${\varepsilon _k}$. Increasing the number of layers may decrease this gap but may also lead higher running time.

\begin{figure}[htbp]
\centerline{\includegraphics[width=3in]{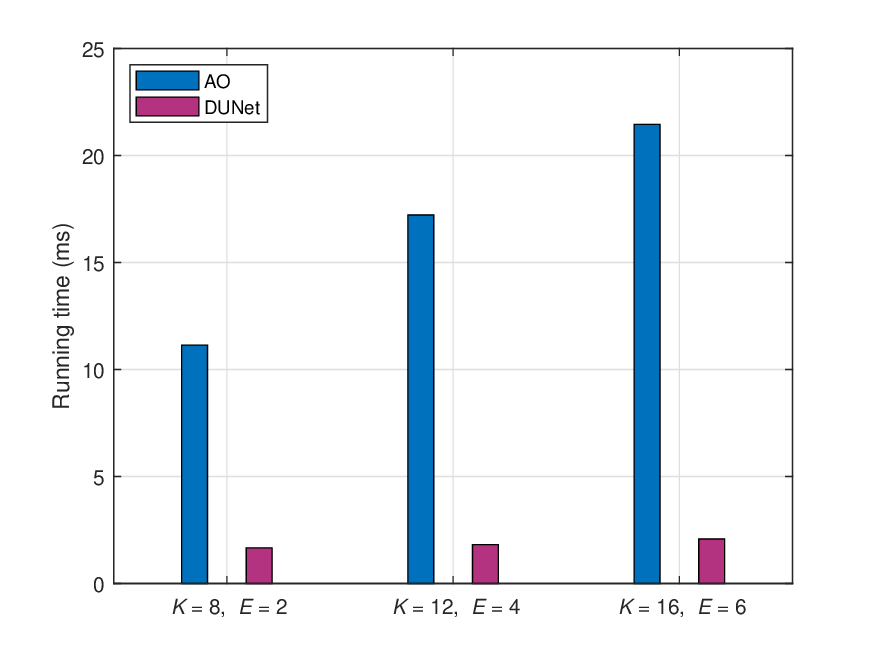}}
\caption{Running time of frameworks versus problem size.}
\label{fig7}
\end{figure}

To compare the performance of AO and DUNet in terms of running time, we plot the results for different number of users while fixing all other parameters as shown in Fig. \ref{fig7}. It is very obvious from the figure that the increasing of the problem size (i.e., $K$ and $E$ ) leads to drastic increase of the running time in case of AO. On the other side, the impact of the increasing of the problem size is marginal for the case of DUNet. This leads to the conclusion that the proposed DUNet is more suitable for real-time application compared with AO.
\section{Conclusion}
In this letter, we proposed a design of a deep-unfolding based framework to handle the beamforming and the phase shift in RIS-empowered secure multi-user communication. First, we formulated a secrecy rate maximization problem; then, we relaxed the problem and designed AO solution. Second, we applied KKT conditions to obtain closed-form solutions for the beamforming and the phase shift. Finally, we transformed the closed-form solutions into deep learning model using deep unfolding mechanism. The proposed DUNet achieves better performance than AO in terms of inference speed and closer to AO in terms of accuracy.

\ifCLASSOPTIONcaptionsoff
  \newpage
\fi


\begin{thebibliography}{1}
\bibitem{md1} M. Poulakis, ``6G's Metamaterials Solution: There's plenty of bandwidth available if we use reconfigurable intelligent surfaces,'' \emph{IEEE Spectrum,} vol. 59, no. 11, pp. 40-45, Nov. 2022.
\bibitem{md2}S. Zhang, M. Li, M. Jian, Y. Zhao and F. Gao, ``AIRIS: Artificial intelligence enhanced signal processing in reconfigurable intelligent surface communications,'' \emph{China Commun.,} vol. 18, no. 7, pp. 158-171, July 2021.
\bibitem{bb1}S. Li, B. Duo, X. Yuan, Y. Liang, and M. D. Renzo, ``Reconfigurable Intelligent Surface Assisted UAV Communication: Joint Trajectory Design and Passive Beamforming,'' \emph{IEEE Wireless Commun. Lett.,} vol. 9, no. 5, pp. 716-720, 2020.
\bibitem{bb2}B. Feng, J. Gao, Y. Wu, W. Zhang, X. G. Xia, and C. Xiao, ``Optimization Techniques in Reconfigurable Intelligent Surface Aided Networks,'' \emph{IEEE Wireless Commun.,} vol. 28, no. 6, pp. 87-93, 2021.
\bibitem{bb3}B. Xiong, Z. Zhang, H. Jiang, J. Zhang, L. Wu, and J. Dang, ``A 3D Non-Stationary MIMO Channel Model for Reconfigurable Intelligent Surface Auxiliary UAV-to-Ground mmWave Communications,'' \emph{IEEE Trans. Wireless Commun.,} pp. 1-1, 2022.
\bibitem{bb4}K. Tian, B. Duo, S. Li, Y. Zuo and X. Yuan, ``Hybrid Uplink and Downlink Transmissions for Full-Duplex UAV Communication With RIS,'' \emph{IEEE Wireless Commun. Lett.,} vol. 11, no. 4, pp. 866-870, April 2022.
\bibitem{bb6}L. Wei, K. Wang, C. Pan and M. Elkashlan, ``Secrecy Performance Analysis of RIS-Aided Communication System With Randomly Flying Eavesdroppers,'' \emph{IEEE Wireless Commun. Lett.,} vol. 11, no. 10, pp. 2240-2244, Oct. 2022.

\bibitem{md3} A. B. M. Adam, M. S. A. Muthanna, A. Muthanna, T. N. Nguyen and A. A. A. El-Latif, ``Toward Smart Traffic Management With 3D Placement Optimization in UAV-Assisted NOMA IIoT Networks,'' \emph{IEEE Trans. Intell. Transp. Syst.,} vol. 24, no. 12, pp. 15448-15458, Dec. 2023.
\bibitem{md4}A. B. M. Adam, L. Lei, S. Chatzinotas and N. U. R. Junejo, ``Deep Convolutional Self-Attention Network for Energy-Efficient Power Control in NOMA Networks,'' \emph{IEEE Trans. Veh. Technol.,} vol. 71, no. 5, pp. 5540-5545, May 2022.
\bibitem{md6}A. B. M. Adam, Z. Wang, X. Wan, Y. Xu and B. Duo, ``Energy-Efficient Power Allocation in Downlink Multi-Cell Multi-Carrier NOMA: Special Deep Neural Network Framework,'' \emph{IEEE Trans. Cogn. Commun. Netw.,} vol. 8, no. 4, pp. 1770-1783, Dec. 2022.

\bibitem{md7}R. Zhong, Y. Liu, X. Mu, Y. Chen and L. Song, ``AI Empowered RIS-Assisted NOMA Networks: Deep Learning or Reinforcement Learning?,'' \emph{IEEE J. Sel. Areas Commun.,} vol. 40, no. 1, pp. 182-196, Jan. 2022.


\bibitem{md8}J. Liu and M. D. Renzo, ``Data-driven and Model-driven Deep Learning Detection for RIS-aided Spatial Modulation,'' \emph{in Proc. IEEE 4th 5G World Forum (5GWF),} Montreal, QC, Canada, 2021, pp. 88-92.


\bibitem{md9}W. Xu, J. An, Y. Xu, C. Huang, L. Gan and C. Yuen, ``Time-Varying Channel Prediction for RIS-Assisted MU-MISO Networks via Deep Learning,'' \emph{IEEE Trans. Cogn. Commun. Netw.,} vol. 8, no. 4, pp. 1802-1815, Dec. 2022.


\bibitem{abm1} Z. Peng, Z. Chen, C. Pan, G. Zhou and H. Ren, ``Robust Transmission Design for RIS-Aided Communications With Both Transceiver Hardware Impairments and Imperfect CSI,'' \emph{IEEE Wireless Commun. Lett.,} vol. 11, no. 3, pp. 528-532, March 2022.
\bibitem{mod211} Z. Li, Q. Lin, Y. C. Wu, D. W. K. Ng and A. Nallanathan, ``Enhancing Physical Layer Security with RIS under Multi-Antenna Eavesdroppers and Spatially Correlated Channel Uncertainties,'' \emph{IEEE Trans. Commun.,} doi: 10.1109/TCOMM.2023.3333919.
\bibitem{bb7}D. W. K. Ng, E. S. Lo and R. Schober, ``Robust Beamforming for Secure Communication in Systems With Wireless Information and Power Transfer,'' \emph{IEEE Trans. Wireless Commun.,} vol. 13, no. 8, pp. 4599-4615, Aug. 2014.
\bibitem{md10} A. B. M. Adam, X. Wan, M. A. M. Elhassan, M. S. A. Muthanna; A. Muthanna, N. Kumar, M. Guizani, ``Intelligent and Robust UAV-Aided Multiuser RIS Communication Technique With Jittering UAV and Imperfect Hardware Constraints,'' \emph{IEEE Trans. Veh. Technol.,} vol. 72, no. 8, pp. 10737-10753, Aug. 2023.
\bibitem{bb8}S. Khobahi and M. Soltanalian, ``Model-Based Deep Learning for One-Bit Compressive Sensing,'' \emph{IEEE Trans. Signal Process.,} vol. 68, pp. 5292-5307, 2020.



\end{thebibliography}
\end{document}